\documentstyle[11pt,newpasp,twoside]{article}
\input epsf

\begin{document}
\title{Spectral Line Advanced Topics}
\author{John M. Dickey}
\affil{University of Minnesota, Department of Astronomy,\\
116 Church St. SE, Minneapolis, MN  55455 USA}

\begin{abstract}

This chapter is in three parts.  The first gives the fundamental
equations of level populations and radiative transfer which
govern spectral line emission and absorption.  
Lines in the cm-wave band have similarities which allow
us to simplify the equations.  The further simplification
of a two level system (a good approximation for H{\tt I})
gives the familiar formulae for the 21-cm line.  

The second part of the lesson deals with mapping, and 
the combination of single dish and interferometer
data.  We consider the effect of gridding
single dish data using a convolving function.  Mapping
speeds and techniques are discussed.

The third part discusses the spectral line cube and its
moments.  The use of the first moments to determine the
dynamical structure of a disk is discussed.  The need for
windowing in the cube before computing moments is
justified.
 
\end{abstract}

\section{Spectral Line Basics :  Radiative Transfer}

Spectral lines below 10 GHz come from transitions
between quantum levels with miniscule energy separations; this
is necessary if the photon energy is to be less than
4$\times 10^{-5}$ ev, so that
its frequency will be in this range.  Atomic hydrogen
has two ranges where the energy levels are so 
closely spaced :  at very high quantum numbers ($n \ga 80$)
and in the ground state itself due to the hyperfine
splitting.  Several molecules also have transitions in
this range, generally caused by similar hyperfine splitting
of energy levels due to the magnetic moment of the nucleus, as in
OH and H$_2$CO.  
The more important spectral lines with frequencies below 10 GHz 
are listed on table 1.  

Recombination lines are generated as
an electron cascades down the energy levels of an atom; at
cm-waves the transitions between levels with 
high quantum numbers are important.
These make a comb of lines whose frequencies are given by the
Rydberg formula, 
\begin{equation} \nu_{ul} \ = \ Z
\ \frac{R_0}{1 + \frac{m_e}{M_{amu}}} \times
\left( \frac{1}{l^2} \ -\ \frac{1}{u^2} \right) \end{equation}
\noindent where   $l$ and $u$ are the quantum numbers of the lower and
upper states of the transition, $Z$ is the screened nuclear charge (i.e.
the number of protons minus the number of other electrons), $M_{amu}$ is
the atomic mass in amu (1.007825 for H),
and $m_e$ is the electron mass.  The Rydberg constant, $R_0$ is 3.289842 $\times$
10$^{15}$ Hz (Rohlfs and Wilson, 1996, p. 316).  
Recombination lines always come in groups, with
the hydrogen line slightly lower in energy than the
corresponding He line, and the corresponding C (and heavier
element) lines piling up at frequencies just slightly
higher than He.  A few examples of recombination lines
are included on the lower half of table 1.

\begin{table}
\caption{Some Spectral Lines with Frequencies Below 10 GHz}
\begin{tabular}{|lll|}
\hline
Transition & Rest Frequency (GHz) & A Recent Reference\\
\hline
H{\tt I} & 1.4204058 & Gibson et al. 2000\\
OH & 1.665402, 1.667385 & Liszt and Lucas 1996\\
& 1.61223, 1.720559 & Lewis et al. 2001 \\
OH & 4.765562, 4.660242 & Szymczak et al. 2000\\
CH & 3.2638, 3.3355, 3.3492 & Magnani and Onello 1993 \\
H$_2$CO & 4.82966 & Pauls et al. 1996 \\
OH & 6.035092 & Caswell 1997 \\
CH$_3$OH & 6.668518 & Phillips et al. 1997 \\
$^3$He & 8.665 & Bania et al. 1997\\
\hline
&&\\
& A Few Recombination Lines&\\
H92$\alpha$ & 8.309382 & Lang et al. 2001 \\
H109$\alpha$ & 5.008923  & Peck et al. 1997 \\
H271$\alpha$ & 0.3285959 & Roshi and Anantharamaiah 1997\\
C271$\alpha$ & 0.3287597 & Roshi and Anantharamaiah 1997\\
\hline
\end{tabular}
\end{table}

\subsection{Level Populations}

Line intensities are determined by the populations of the two quantum
levels of the transition, which set the amount of emission and absorption
at each point along the line of sight.  
Whether or not the excitations are in thermal equilibrium
with the kinetic temperature of the gas, we can describe the ratio
of the populations of the two levels using the Boltzmann equation with
some temperature $T_{ex}$, as :
\begin{equation} \frac{n_{u}}{n_{l}} \ \ = \ \ 
\frac{b_{u}}{b_{l}} \ \frac{g_{u}}{g_{l}} \
e^{\frac{-h \nu}{kT_{ex}}} \end{equation}
where $n_u$ and $n_l$ are the level populations of the upper and lower levels,
$g_u$ and $g_l$ are the statistical weights of the two levels,
$h \nu$ is the energy of the photon, $k$ is Boltzmann's constant, 
and $T_{ex}$ is a temperature which is equal to the kinetic
temperature in the special case where the collision rate
is high enough to thermalize the transition (Osterbrock 1989, section 3.5).  

At cm-waves the photon energies are so low that
we can usually make the approximation $\frac{h \nu}{k T_{ex}} \ll 1$;
in that case we can expand the Boltzmann equation in a Taylor series :
\begin{equation} \frac{n_u}{n_l} \  =  \ \frac{g_u}{g_l}\ \times \left( 1\  -
\ \frac{h \nu}{k T_{ex}} 
\ + \cdots \right)\end{equation}
For a two level system, i.e. one in which all the atoms are 
in one or the other of two levels, so that $n_u + n_l = n$, 
the first order term of the expansion makes the
number in the upper level, $n_u$, just proportional to the total 
density, i.e.
\begin{equation} n_u \ = \ n \times \frac{n_u}{n_u + n_l} \ = \ n \times \frac{
\frac{n_u}{n_l}}{1 + \frac{n_u}{n_l}} \ \simeq
\ n \times \frac{g_u}{g_u+g_l} \end{equation}
this gives $n_u  \simeq \frac{3}{4} \times n $
for the hyperfine-spit levels of the H{\tt I} ground state.
The two level approximation is valid for atomic hydrogen when all
the atoms are in the ground state, as in the cool phases of the
interstellar medium.

\subsection{Emission and Absorption}

The emission and absorption coefficients can be derived once we know the
level populations.  Generally the emission
coefficient, $j_{\nu}$, is given by : 
\begin{equation} \int\ j_{\nu} \ d\nu \ =
\ \frac{n_{u} \ A_{ul} \ h \nu}{4 \pi} \end{equation}
where $A_{ul}$ is the Einstein coefficient for spontaneous emission
for the transition from level $u$ to $l$.
The frequency integrals are taken over the line profile 
(the subscript $\nu$ indicates that this is a function of frequency).
The absorption coefficient is given by 
\begin{equation} \int \ I_{\nu} \ \kappa_{\nu} \ d\nu \ = \ h \nu \left(
n_{l} B_{lu}\ - \ n_{u} B_{ul} \right) \frac{I_{\nu}}{c} \end{equation}
where $I_{\nu}$ is the radiation intensity and
$B_{lu}$ and  $B_{ul}$ are the Einstein coefficients which
describe the probability of absorption and stimulated emission
of a photon, respectively.  These are in the ratio 
\begin{equation} B_{lu} \ = \ \frac{g_u}{g_l} B_{ul} \end{equation}
(Spitzer 1977, section 3.4).

Note that the units of the emission and absorption coefficients
are different.
For $j_{\nu}$ the units are erg cm$^{-3}$ sec$^{-1}$ Hz$^{-1}$
sterrad$^{-1}$, which is just the energy in the radiation coming
from a unit volume of gas per second, going into a unit bandwidth of
the emission spectrum and into a unit solid angle in direction.
The units of $\kappa_{\nu}$ are the same, divided by I$_{\nu}$, since
the absorption is always proportional to the
intensity of the incident radiation field itself; this works out
to give $\kappa_{\nu}$ simply units of cm$^{-1}$.  Thus the combined
effect of the emission and absorption of a differential volume element,
$ds$, somewhere along the line of sight
is to change the radiation intensity by $dI$ where 
\begin{equation} d I_{\nu} \ = \ j_{\nu}\ ds \  
\ - \ \kappa_{\nu} \ \ I_{\nu} \ ds \end{equation}

The ratio of the emission and absorption
coefficients is the Plank source term :
\begin{equation}
  {\mathcal{B}}_{\nu}\ \ = \ \  \frac{j_{\nu}}{\kappa_{\nu}}
\end{equation}
\begin{equation}
 \ \ \ \ \ \ \ \  = \ \ \frac{2 h \nu^{3}}{c^{2}} \ \frac{1}
{e^{h \nu/k T} \ - 1}
\end{equation}
\begin{equation}
 \ \ \ \ \ \ \ \ \ \simeq \ \frac{2 k T}{\lambda^2}
\end{equation}
where the approximation is the Rayleigh-Jeans law which defines the
brightness temperature :
\begin{equation}
 T_B\ \  = \ \  \frac{{\mathcal{B}}_{\nu}\ c^2}{2\ k\ T\ \nu^2}
\end{equation}

For a two level system with energy separation giving a cm-wave line, the
emission coefficient is proportional to the density, using equation 3 in 
equation 5 gives: 
\begin{equation} \int j_{\nu} d\nu \ = \  \frac{g_u}{g_u + g_l} \  
\frac{A_{ul} \ h \nu}{4 \pi} \times n \end{equation}
For the absorption coefficient we must include the second term in the
Taylor expansion in equation 3, since to first order the number of photons
absorbed is just cancelled by the extra number produced through
stimulated emission.   Thus we get 
\begin{equation} \int \kappa_{\nu} d\nu = \frac{h \nu\ B_{ul}}{c\  g_l} \left(
\ g_u n_l \ - \ g_l n_u \right) \end{equation}
\begin{equation} \ \ \ \ \ = \  \frac{h \nu\ B_{ul}\ n_l}{c\  g_l}
 \left( g_u \ - \ 
g_l \frac{n_u}{n_l} \right) \end{equation}
\begin{equation} \ \ \ \ = \  \frac{h \nu\ B_{ul}\ n_l}{c\  g_l}
\left\{ g_u \ - \
g_l \left[ \frac{g_u}{g_l} \left( 1-\frac{h \nu}{k T_{ex}} \right)
\right] \right\} \end{equation}
\begin{equation} \ \ \ \ = \  \frac{(h \nu)^2\ B_{ul}\  n}{c\ k\ T_{ex}}
\ \left( \frac{g_u}{g_u + g_l} \right)
\end{equation}

\subsection{Radiative Transfer}

When we integrate along the line of sight to determine what the telescope
sees, the emission integral gives 
\begin{equation} 
I_{\nu} \  = \ \int j_{\nu}\ dx \ = \ \frac{A_{ul}\ h \nu}{4 \pi} 
\frac{g_u}{g_u + g_l} \int n\ dx 
\end{equation}
or in brightness temperature units with Doppler velocity in place of frequency
\begin{equation} 
\int T_B(v)\ dv \ = \ C_0 \times N
\end{equation}
where C$_0$ = 5.485 $\times 10^{-19}$ K km s$^{-1}$ cm$^{2}$
for the 21-cm line.  Note that we have to
take the velocity integral as well as the line of sight integral to
determine the total column density, $N$, because the radial motions
of the atoms give them slight Doppler shifts which generate a line
profile in velocity or frequency.  
In the case of an unresolved object such as a distant galaxy, 
equation 19 gives the HI mass by 
\begin{equation}
\frac{M_H}{m_{\sun}} \ = \ 2.3 \times 10^5 \ \left( \frac{d}{Mpc} \right)^2
\ \frac{\int S_v \ dv}{Jy \ km s^{-1}}
\end{equation}
The corresponding integral of the
absorption coefficient along the line of sight gives the optical depth :
\begin{equation} 
\tau_{\nu} \ = \ \int \kappa_{\nu} \ dx \ = \ \frac{(h \nu)^2\ B_{ul}\  n}
{c\ k\ T_{ex}} \frac{g_u}{g_u + g_l} 
\ \int \frac{n}{T} \  dx  
\end{equation}
or for the velocity integral 
\begin{equation} 
\int \tau_{v} dv \ = \ C_0 \ \frac{N}{T_{harm}} 
\end{equation}
where again we replace the frequency integral with an integral over
Doppler velocity, and $C_0$ for the 21-cm line has the same value
as above.  The $T_{harm}$ in equation 21 is the harmonic mean
temperature,
\begin{equation}
T_{harm} \ \equiv \ \frac{\int n \ dx} { \int \frac{n}{T_{ex}}\ dx}
\end{equation}
This is only equal to the excitation temperature if the gas is isothermal along
the line of sight.

\epsfxsize=4.5in
%\hspace{.5in} \epsfbox{Bozofig1.eps}
\epsfbox{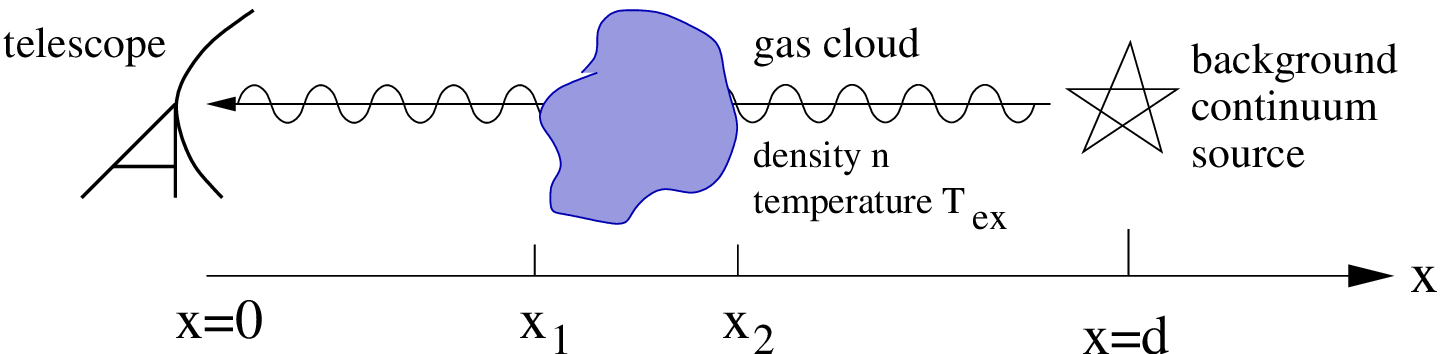}

\noindent {\small Figure 1.  The geometry for the
radiative transfer calculation in equations 29 - 31.}

Once we have determined the emission and absorption coefficients we can
use radiative transfer to calculate how the radiation intensity builds up
through propagation along the line of sight, as shown
in figure 1.  Dividing equation 8 by $\kappa_{\nu}$ gives 
\begin{equation}
 \frac{d I_{\nu}}{d\tau_{\nu}}
\ = \  -\ \frac{j_{\nu}}{\kappa_{\nu}} \ + \ I_{\nu} 
\end{equation}
since $d \tau = \kappa_{\nu}\ dx$ and $dx = -ds$, i.e. $x$ increases away
from the telescope, opposite the direction of propagation, $ds$.
We can integrate this along the line of sight, 
$x=0$ to distance $d$ which translates to $\tau = 0$ to $\tau(d)$, 
after multiplying both sides by $e^{-\tau}$, i.e.
\begin{equation}
 e^{-\tau_{\nu}} \ \frac{d I_{\nu}}{d\tau_{\nu}}  \ - \
I_{\nu} e^{-\tau_{\nu}} \ = \ - \frac{j_{\nu}}{\kappa_{\nu}} \
e^{-\tau_{\nu}}  
\end{equation} or
\begin{equation}
 \frac{d}{d\tau_{\nu}} \left( e^{-\tau_{\nu}} \ I_{\nu} \right)\
 \ = \ - \frac{j_{\nu}}{\kappa_{\nu}} \ e^{-\tau_{\nu}} 
\end{equation}
\begin{equation}
 \int_{I(0)}^{I(d)} \ d\left( e^{-\tau} I_{\nu} \right) \ = \ - \
\int_{0}^{\tau(d)} \ \frac{j_{\nu}}{\kappa_{\nu}} \ e^{-\tau} \ d\tau 
\end{equation}
\begin{equation}
 e^{-\tau(d)} I(d) \ - \ e^{-\tau(0)} I(0) \ = \ - \
\int_{0}^{\tau(d)} \ \frac{j_{\nu}}{\kappa_{\nu}} \ e^{-\tau} \ d\tau 
\end{equation}
Since we are starting at the telescope, $\tau(0)=0$, so
\begin{equation} I(0) = e^{-\tau(d)} I(d) \ + \
\int_{0}^{\tau(d)} \ \frac{j_{\nu}}{\kappa_{\nu}} \ e^{-\tau} \ d\tau 
\end{equation}
A simple case is that of a single cloud with a continuum background
source, so that ${\mathcal{B}}_{\nu}(x) = {\mathcal{B}}_o$
inside the cloud (distance $x_1$ to $x_2$) 
and zero everywhere else, and $I(d) \equiv  I_{bkg}$.  Then 
\begin{equation}
\tau \ \ = \ \ \ \int_{x_1}^{x_2}\ \kappa \  dx 
 \ \ \ \ \ \ \ = \ \kappa (x_2 - x_1)
\ = \ C_0 \times \frac{N}{T_{ex}} 
\end{equation}
\noindent where $N = n (x_2 - x_1)$ is  the column density through the
cloud, and the velocity integral has been taken on both sides.
Then the received intensity of the radiation is 
\begin{equation}
I_0 \ = \ {\mathcal{B}}_c \ \left( 1 - e^{-\tau} \right) \ + \ I_{bkg}
\ e^{-\tau} 
\end{equation}
where ${\mathcal{B}}_c$  is the Plank function for the excitation 
temperature of the gas in the cloud.  In brightness temperature notation 
this is
\begin{equation}
T_B \ = \  T_{ex} \left(1\ -\ e^{-\tau} \right) + T_{bkg} \ e^{-\tau}.
\end{equation}

\epsfxsize=4.5in
%\hspace{.5in} \epsfbox{Bozofig2.eps}
\epsfbox{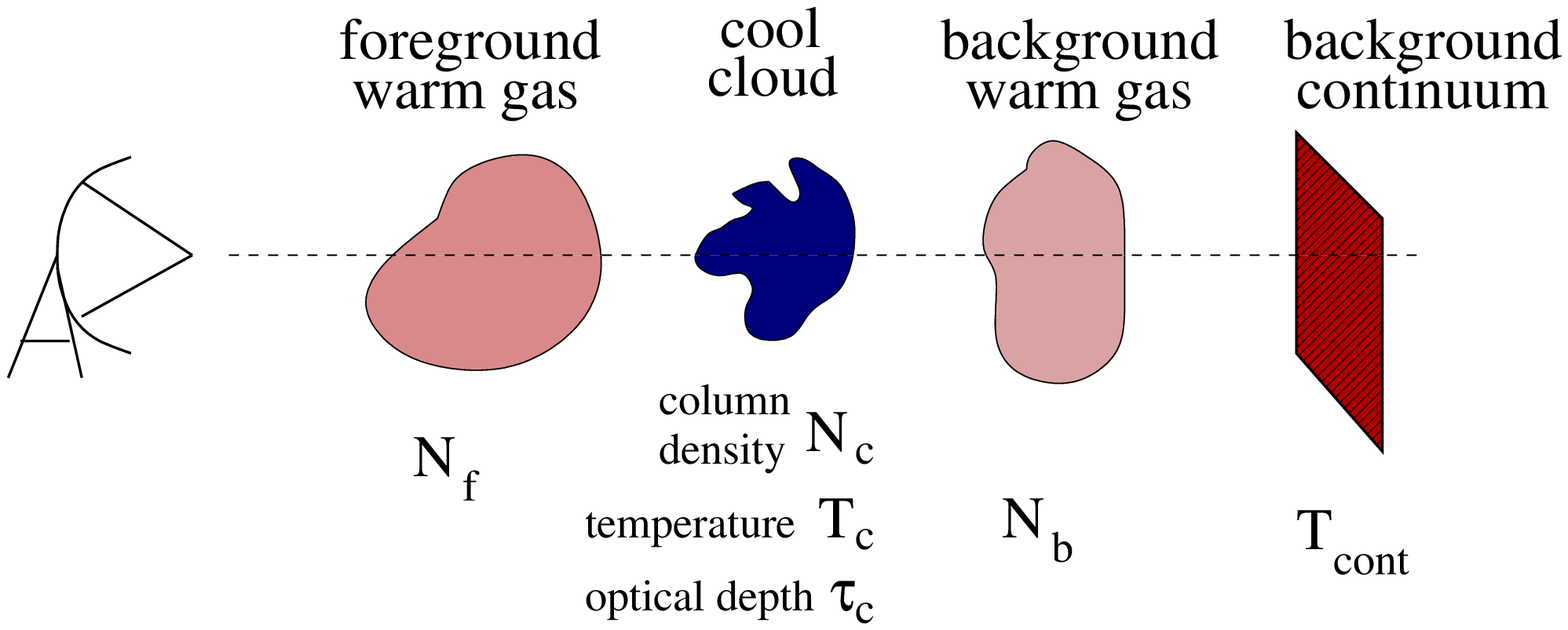}

\noindent {\small Figure 2.  The geometry for the
radiative transfer calculation in equation 32. }

A more complicated case is shown on figure 2,
with optically thin foreground and background
clouds of gas (column densities $N_f$ and $N_b$)
that are warm enough to have negligable optical depth,
plus a cool, absorbing cloud with temperature $T_c$ and column density
$N_c$, plus background continuum, $T_{cont}$.
This case gives 
\[ T_B (v) \ \ = \ \ \frac{C_0 \ N_f}{\sqrt{2 \pi}\ \sigma_f}
\ e^{\frac{(v-v_f)^2}
{2 \sigma_f^2}} \ + \ T_c\ 
\ \left( 1 - e^{- \tau (v)} \right) \ + \]
\begin{equation}
\ \ \ \ \ \ + \ \left( T_{cont} \ + \ \frac{C_0 \ N_b}{\sqrt{2 \pi}\ \sigma_b}
\ e^{\frac{(v-v_b)^2}
{2 \sigma_b^2}} \right)\  \ e^{- \tau (v)} 
\end{equation}
Here we assume that the shapes of the profiles are Gaussians with center
velocities and dispersions ($v_f$, $\sigma_f$), ($v_c, \sigma_c$), and
($v_b$, $\sigma_b$) for the foreground, cloud, and background gas, respectively.

\subsection{Velocity Profiles}

In most areas of astronomy we use the optical version of the Doppler
shift formula, 
\begin{equation}
v_{radial} \ \equiv \  c z \ = \ c \times 
\frac{\Delta \lambda}{\lambda_r}
\end{equation}
where $\Delta \lambda$ is the difference between the observed wavelength and
the rest wavelength, $\lambda_r$.  This defines the redshift, $z$,
without any relativistic corrections, so that $z$ can be greater than
one.  A useful formula to remember is that {\bf the velocity
range corresponding
to one MHz of bandwidth is just equal to the wavelength in mm}.  
For the $\lambda$21.1 cm line of HI this means that 1 MHz corresponds to 
211 km s$^{-1}$, or 1 km s$^{-1}$ is $\frac{1000}{211}$ kHz or 4.73 kHz.

For a Maxwellian distribution of velocities corresponding to a thermal
gas with temperature $T_{kin}$, the line profile is Gaussian, and
so $j_{\nu}$, $\kappa_{\nu}$, and $I_{\nu}$ all have frequency or
Doppler velocity profiles following
\begin{equation}
I_{v} \  = \ I_{0} \times e^{\left[ -\  \frac{\left( v-v_0 \right)^2}
{2 \sigma_v ^2 } \right]}
\end{equation}
with 
\begin{equation}
\sigma _v = \sqrt{\frac{2 k T }{ M_{amu}}}.
\end{equation}  This reduces to
\begin{equation}
\sigma _v = 0.91 km s^{-1} \times \sqrt{ \frac{T}{100K} \times
\frac{1}{M_{amu}}}
\end{equation}
for hydrogen, $\sigma _v \ \simeq$  1 km s$^{-1}$ for gas at
120 K temperature,
$ \sigma _v \ \simeq$10 km s$^{-1}$ for gas at 12,000 K , and
$ \sigma _v \ \simeq$100 km s$^{-1}$ for
gas at 1.2 million K.

\section{Mapping Basics :  The uv Plane}

Whether we are observing with a single dish telescope, an aperture
synthesis telescope, or a combination of the
two, it is crucial to keep in mind the two Fourier conjugate
representations of the telescope beam and the sky brightness.  
These are shown on figure 3, with the uv plane functions on the left,
and their conjugate functions on the plane of the sky on the right.
Since the aperture plane is the transform of the image plane, the 
telescope beam is the Fourier transform of the aperture illumination
in both cases. But there is a fundamental difference between a 
single dish telescope (and/or an adding array interferometer) and 
a multiplying or correlation interferometer (aperture synthesis
telescope).  In the single dish case (shaded on the lower right of 
figure 3) it is the autocorrelation
function of the illumination pattern which matters; the transform of
this function is the beam power pattern.  For the multiplying 
interferometer (shaded on the upper left of figure 3)
the synthesized beam (dirty beam) is the transform
of the illumination pattern itself.  Thus there are negative 
sidelobes in the latter case, but not the former, since the autocorrelation
function is always symmetric, so that its transform is real
and positive definite. 

\epsfxsize=4in
\hspace{.5in} \epsfbox{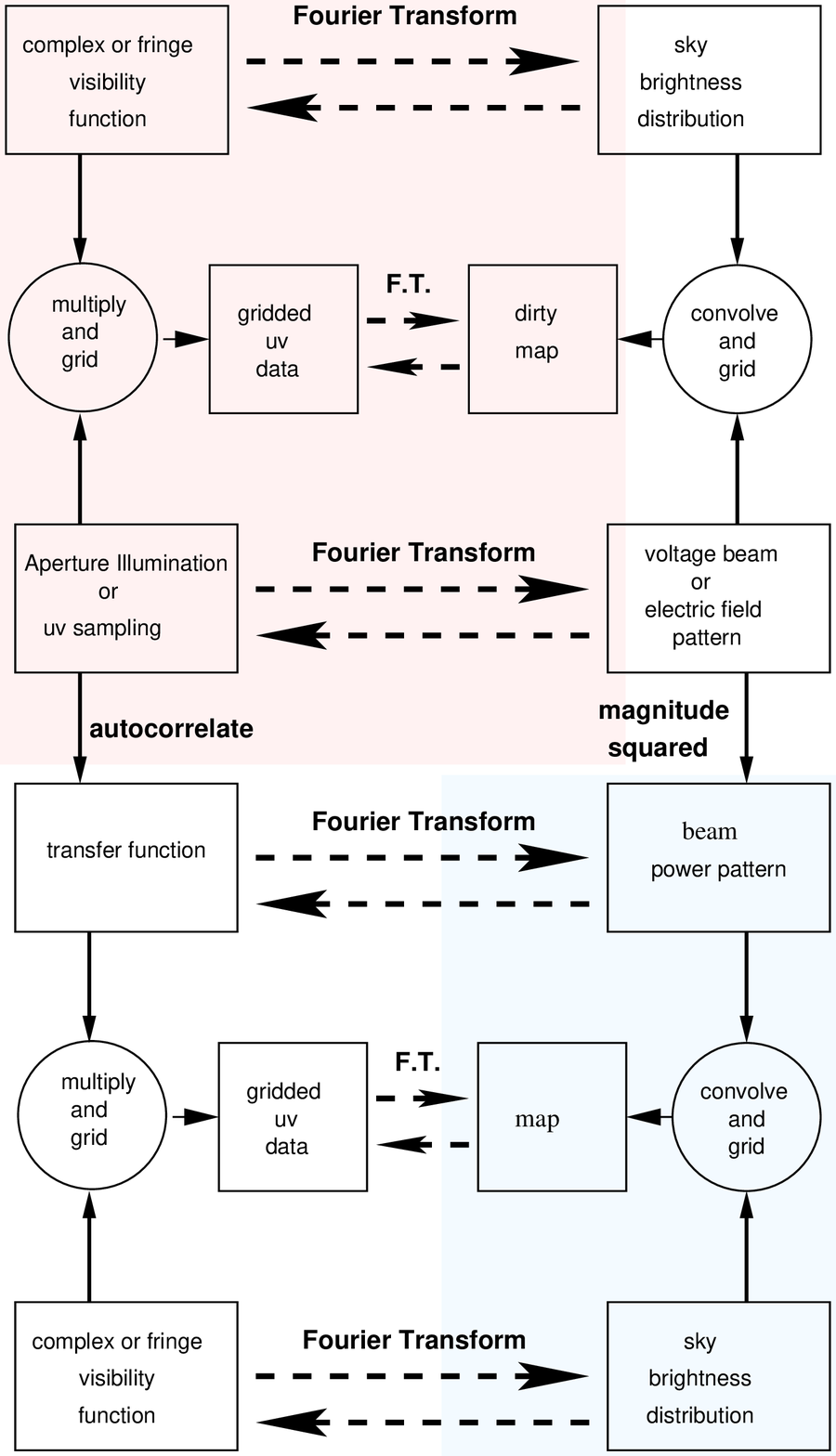}

\noindent {\small Figure 3.  The relationship between the 
instrumental response and the sky brightness which lead
to the measured map or image of the sky for two types of
telescopes, aperture synthesis telescopes,
i.e. correlation interferometers, (upper shaded area) and single
dish telescopes or adding interferometers (lower shaded
area).  The left side is the $uv$ plane, the right side is
the $lm$ or sky plane.}

\subsection{Gridding}

An important step in the processing of either interferometer
or single dish data is the gridding process.  The Nyquist relation
requires that the pixels in the map plane should be small enough that
the FWHM of the beam is at least 2.4 pixels across, 
preferably 2.8 or more.  If the pixels are larger than this
relative to the beam width, then
the higher spatial frequency information is lost due to 
undersampling.  This is controlled in the aperture synthesis
case by the gridding of the interferometer data on the uv plane
before the transform back to the sky plane.  When mapping with
a single dish a similar gridding step is needed to make 
a map out of data taken either in separate pointings or 
``on-the-fly'' while the telescope is driving.  Even if the
single dish observations are carefully taken in a lattice
of pointings, it is still helpful to regrid the data using
a convolving function which controls the behavior of the
transform of the map.  For example, noise in the spectra
from the individual pointings can appear to have very high
spatial frequencies, which alias on the uv plane back to
low spatial frequencies unless they are suppressed by
a convolving function in the gridding process.  Thus the
gridded map is :
\begin{equation}
T(x,y)\  = \ \frac{\Sigma _i  T_i(l,m) f\left[ (l-x), (m-y) 
\right] } {\Sigma _i  U_i(l,m) f\left[ (l-x), (m-y)
\right] }
\end{equation}
where the individual observations $T_i$ are taken at 
positions $(l,m)$
and the regular grid of the map pixels is given by $(x,y)$.
$f(x,y)$ is the convolving function, and $U(x,y)$ is a normalization
function which is one at the positions of the observations.
The division is suppressed for positions $(x,y)$ so far from
any observation that the denominator is less than some minimum
threshold, typically between 0.2 and 0.05.  A Gaussian is often
used for $f$ in the sky plane, in the uv plane it is
better to use a exponential times sinc because of its aliasing
suppression properties.  If a Gaussian is used, the gridded
map has resolution given by 
\begin{equation}
FWHM_{gridded}^2 \ = \ \sqrt{FWHM_{original}^2 + FWHM_f^2}
\end{equation}

The normalization illustrated
in equation 37 by $U$ has many variations, especially
in aperture synthesis applications.  The easiest to understand
are ``uniform weighting'', in which each $uv$ cell has the
same weight regardless of the number of samples which
contribute to it, and ``natural weighting'', in which each
sample has the same weight, so that regions which are heavily
sampled are weighted more highly than sparsely sampled regions.
``Robust weighting'' achieves some of the best features of
both strategies.  In any event, the gridding should provide pixels
in the sky plane spaced by less than 1/2.4 times the FWHM of
the telescope beam.

\subsection{Mapping Speeds}

For single dish mapping the speed depends on the desired 
sensitivity of the spectra.  Using the radiometer equation :
\begin{equation}
\sigma_T = \frac{\sqrt{2} \ T_{sys}}{\sqrt{\delta \nu \ T_{int}}}
\end{equation}
with $\delta \nu$ the channel bandwidth, $T_{int}$ the integration
time, and $T_{sys}$ the system temperature, we get the rms noise,
$\sigma_T$.  The factor of $\sqrt{2}$ in the numerator 
is usually needed to account for calibration, 
e.g. frequency switching.
For the example of on-the-fly mapping, we should
drive the telescope and read out the spectra so that one
spectrum is read in the time it takes the telescope to move
by less than 1/2.4 times the beamwidth, and so that this time
provides $\sigma_T$ less than 1/5 times the weakest features we
hope to measure in the spectra after gridding.  For Arecibo at 21-cm, this
means we should read out spectra spaced by no more than about
1.2\arcmin.  For channel bandwidth $\delta \nu$
of 0.5 km s$^{-1}$ or 2.4 kHz we would
need integration time per spectrum of 75 seconds to achieve
rms noise of 100 mK, assuming system temperature of 30 K.
Thus we can cover one square degree with 2600 spectra representing 
total integration time of 54 hours (plus overhead for telescope
motion and calibration).  So single dish mapping can be a 
slow process, even at such a modest sensitivity as this.
The most common response to this is to undersample the
beam, spacing the spectra further apart than the Nyquist
condition requires.  This compromises the quality of the
map, particularly if it is to be combined with interferometer
data.  It also causes the gridded map to have an uneven sensitivity
function, which is given by the noise divided by the normalization in 
equation 37, i.e.
\begin{equation}
S(x,y)\  = \ \frac{\sigma_T }
 {\sqrt{\Sigma _i U_i(l,m) \ f\left[ (l-x), (m-y) \right] }}
\end{equation}

\subsection{Multibeam Surveys and Mosaicing}

A very effective response to this problem is to build a multibeam
receiver.  An example is the Parkes
21-cm multibeam, which has been in use for more than four years on 
the 64m telescope of the Australia Telescope National Facility.
%(figure 4)
Using just the seven inner beams a team of astronomers
led by Naomi McClure-Griffiths have used this instrument to map the
Southern Milky Way at 21-cm (McClure-Griffiths et al. 2000, 2001).
This project is similar to a Northern Milky Way survey, the Canadian
Galactic Plane Survey (CGPS, English et al. 
1998, Normandeau et al. 1996).  Both surveys combine single dish
and aperture synthesis data to achieve high dynamic range, 
good resolution, and most of all uniform sensitivity to all angular
scales from $\sim$1\arcmin \ to many degrees.

Figure 4 indicates the telescope time required to survey a given area
to a surface brightness sensitivity of 1 K (rms) in a velocity channel
with width 0.8 km s$^{-1}$.  The crosses mark single dish telescopes,
while interferometers are marked with crescents which illustrate how
tapering (weighting down the longest baselines) can improve the brightness
sensitivity.  Only a little tapering helps; tapering to beamwidths 
larger than about two times the untapered value decreases the 
brightness sensitivity because so much data is weighted down by
the tapering function.   In boxes on figure 4 are shown three 
future instruments.  The Arecibo multibeam illustrates with the
dashed arrow how much improvement in survey efficiency could be
achieved if a seven beam system were installed on the AO telescope.
The E array is a possible enhancement to the VLA which would 
concentrate the antennas even more densely than in the D array.
The SKA is the Square Kilometer Array.  The design of this
telescope is not settled yet, but it will provide several 
orders of magnitude improvement in brightness sensitivity and
resolution over any existing cm-wave telescope.

\epsfxsize=5.5in
%\hspace{-.8in} \epsfbox{Fig4.eps}

\vspace{.5in}
\noindent {\small Figure 4.  [{\it See the jpeg file associated with 
this paper.}]
Mapping speeds with various telescopes.  The resolution is
indicated on the horizontal axis, while the number of square degrees
per hour is on the vertical axis.  The single dish telescopes 
(marked with crosses) would never be driven so fast as this figure
suggests.  Driving them more slowly (say, at 1 degree per hour) would
give rms noise, $\sigma_T$, smaller than 1 K in proportion to the
inverse square root of the effective integration time.
}
%\epsfxsize=3.5in
%\hspace{.8in} \epsfbox{multib.eps}

%\noindent {\small Figure 4.  The Parkes multibeam receiver
%being lifted into position in the focal cabin.}

Large area surveys like the Southern Galactic Plane Survey (SGPS) and
the CGPS using interferometer telescopes must make use of the
technique of mosaicing.  This observing strategy moves the 
pointing center frequently (typically every 30 seconds) over a 
raster of positions spaced by roughly the half width (FWHM/2) of the 
primary beam of the interferometer.
These ``snap-shot'' observations are repeated many times to build up
good $uv$ coverage.  This is a more effective use of telescope
time than longer integration times 
per snap-shot, since the earth turns so slowly that most baselines
do not change grid cells on the $uv$ plane for many minutes.
The more profound advantage of combining data from many nearby
pointing centers of the interferometer is that it allows reconstruction
of the shorter $uv$ spacing information which cannot be directly
measured because the antennas would shadow each other on such short
baselines (Ekers and Rots, 1978, Cornwell 1985, Sault et al. 1995).
The effect of mosaicing on the $uv$ plane is to deconstruct the
antennas, and give information for all baselines corresponding to 
the nearest bit of one antenna to the nearest bit of the next.

\subsection{Combining Single Dish and Aperture Synthesis Data}

Combining the data from the two types of telescopes can be
done in a variety of ways.  These were compared quantitatively
using 21-cm data on the Small Magellanic Cloud by Stanimirovi\'c
(1999).  The basic alternatives are to combine the uv data
before mapping, to combine dirty maps and jointly deconvolve
the beam shapes (implemented in the Miriad task {\tt MOSMEM}),
and to deconvolve separately and combine the cleaned maps
(implemented in the Miriad task {\tt IMMERGE}).  The last
has given consistent and trustworthy results in many tests.

However they are combined, the relative calibration
of the two sets of data is crucial.  Generally single dish
data is calibrated either using a standard brightness region
(e.g. Weaver and Williams, 1973) which gives units of K for $T_B$, or 
by observing unresolved continuum sources of known flux density,
which gives units of Jy per beam.  Interferometer maps are always
calibrated using unresolved sources, thus they also generally have
units of Jy per beam.  The two are related by the gain of the
synthesized beam, i.e. 
\begin{equation}
G \ = \ \frac{A_e}{k} \ = \ \frac{\lambda^2}{\Omega_B \ k} 
\end{equation}
where $A_e$ is the effective area of the dish and $\Omega_B$
is the solid angle of the synthesized (clean) beam, i.e.
\begin{equation}
\Omega_B \ = \ 1.13 \ FWHM_1 \times FWHM_2
\end{equation}
where $FWHM_1$ and $FWHM_2$ are the major and minor axes of the 
clean beam.  For these beam widths in arc minutes, the gain is
\begin{equation}
G \ = \ 169\  K\  Jy^{-1} \left( FWHM_1 \times FWHM_2 \right)^{-2}
\end{equation}
This definition of the gain is $\frac{\lambda^2}{4 \pi k}$ times
the standard engineering quantity called the {directive gain}.
The advantage of this definition is that it gives the conversion
between units of Jy per beam and K of brightness temperature.
For the Arecibo telescope at 21-cm $G \simeq 10$ K Jy$^{-1}$,
for the GBT it may be about 1.5 K Jy$^{-1}$.  For the VLA
D-array with clean beam size of 45\arcsec \ the gain is 300 K Jy$^{-1}$.
Using the gain, we can convert from the observed antenna temperature to 
the true flux density of the source, $S$.  In the spectral line
case, the analog of the brightness temperature integral of equation 19
becomes the flux integral ($\int S(v) \ dv \ = \ \frac{\int T_A (v) \ dv}{G}$).
This gives the HI mass 
for the case of an unresolved 21-cm line source such as a distant galaxy, 
\begin{equation}
\frac{M_H}{m_{\sun}} \ = \ 2.3 \times 10^5 \ \left( \frac{d}{Mpc} \right)^2
\ \frac{\int S(v) \ dv}{Jy \ km s^{-1}}
\end{equation}
where $d$ is the distance to the galaxy.

\epsfxsize=4.5in
%\epsfbox{Fig5a.eps}

\epsfxsize=4.5in
%\epsfbox{Fig5b.eps}

\epsfxsize=4.5in
%\epsfbox{Fig5c.eps}

\noindent {\small Figure 5. [{\it See the jpeg file associated with 
this paper.}] Images of part of a Galactic
Supershell at longitude 277, latitude 0, velocity 38.75 km s$^{-1}$. 
The upper panel shows Parkes data only, the middle panel shows
ATCA interferometer data only, and the lower panel shows the
combined map. }

Before the data can be combined the calibration scales of the two 
instruments must be matched carefully.  The only sure way to
do this is by comparing data for the same source {\bf in the
overlap region on the $uv$ plane}, which is the range of spatial
frequencies for which both instruments have good sensitivity.
Such an overlap calibration is implemented in the Miriad task
{\tt IMMERGE}.

Figure 5 shows the combination of single dish and interferometer
maps of the edge of a supershell in the outer Milky Way HI 
(McClure-Griffiths et al. 1999).  The range of $uv$ spacings
contributing to each is manifest in its appearance.  Only by
combining them is the full range of structure clear.

\section{Using Spectral Line Cubes}

The spectral line cube is a three dimensional data structure
made by stacking maps taken at different frequencies, i.e.
Doppler velocities.  The cube can be represented as a movie,
either as a series of images of the sky at different frequencies,
or a series of position-velocity diagrams taken along different
lines on the sky.  These and several other very useful
ways of displaying spectral line cubes are implemented in the
{\tt KARMA} package (Gooch 2000).  It is
important not to confuse the spectral line cube with an
image of the line emission in three spatial dimensions, but 
often there is a velocity gradient along the line of sight
due to the dynamics of the system which allows us to link
some velocities with distances.

Another way to represent the data in the cube is by computing
the velocity moments of the brightness.  The $n$th moment map,
$V^n(x,y)$ is defined by
\begin{equation}
V^n(x,y) \ \equiv \ \int_v T_B(x,y,v) \times (v-v_0)^n \ dv
\end{equation}
where the cube is $T_B(x,y,v)$, $v_0$ is the systemic velocity
or some appropriate zero point for the velocity scale,
and the integral is taken as 
a sum over some or all of the planes of the cube.
The zero-th moment map is simply a map of the integral of
the spectrum, as in equation 19 but taken for every spatial
pixel.  For the 21-cm line this gives column density (as long
as the optical depth is not high).  The first moment
map shows the mean velocity (weighted by intensity) in each
pixel.  Often this gives a good image of the velocity field
of the source.  The second moment map generally shows the linewidth
as a function of position.  Examples of moment maps are
shown on figure 6.

The first and higher order moment maps are very unstable to 
bias in the presence of noise.  To be safe, the cube should 
be blanked to eliminate regions in both space and velocity
which have no signal.  This can be done by hand (the {\tt AIPS}
task {\tt BLANK} is a good interactive tool for this), or 
automatically using a smoothed version of the cube as a 
template for determining where there is and is not signal.
The {\tt AIPS} task {\tt MOMNT} is a good implementation of
this strategy; it sets a threshold well outside the region with
detected signal, so as to include the faint tails of
strong lines, but it excludes spurious noise peaks in
regions of the cube far from any emission.

The first moment map of a spectral line from the disk of a
galaxy is generally used to fit the rotation velocity field.
A good task for doing this is the {\tt GIPSY} task {\tt ROCUR},
which has been implemented on some {\tt AIPS} systems.
The objective is to determine several unknown parameters: 
the position of the center of the rotation pattern, the
inclination (perhaps as a function of radius if the galaxy is
warped) and the position angle
of the major axis (i.e. the line of nodes, perhaps also as a function
of radius), and the rotation curve, i.e. the circular velocity as
a function of radius.  The rotation curve is of critical 
importance for dynamical modelling since, with the assumption
of circular rotation, it shows the distribution
of gravitational force as a function of radius.  Departures
from circular rotation, e.g. in spiral arms, can often 
be analysed based on the first moment map as well.

\epsfxsize=5.5in
%\epsfbox{Fig6.eps}

\noindent {\small Figure 6.  [{\it See the jpeg file associated with 
this paper.}]  Moment maps of M31 made from 
the WSRT synthesis data of Brinks and Shane (1984).
The left figure is the zero moment, showing the H{\tt I}
column density, the center figure is the first moment,
showing the radial velocity field, and the right figure
is the second moment, showing the line width.}

Fitting the parameters of the disk rotation gives a model
velocity field which can be subtracted from the first moment
map, leaving the velocity residual map.  A good fit shows only
noise in the residual map, but if the fit is not perfect, the
residual map shows distinct patterns.  Figure 7 shows some
examples, taken from Warner, Wright, and Baldwin (1973).

\epsfxsize=4.5in
\epsfbox{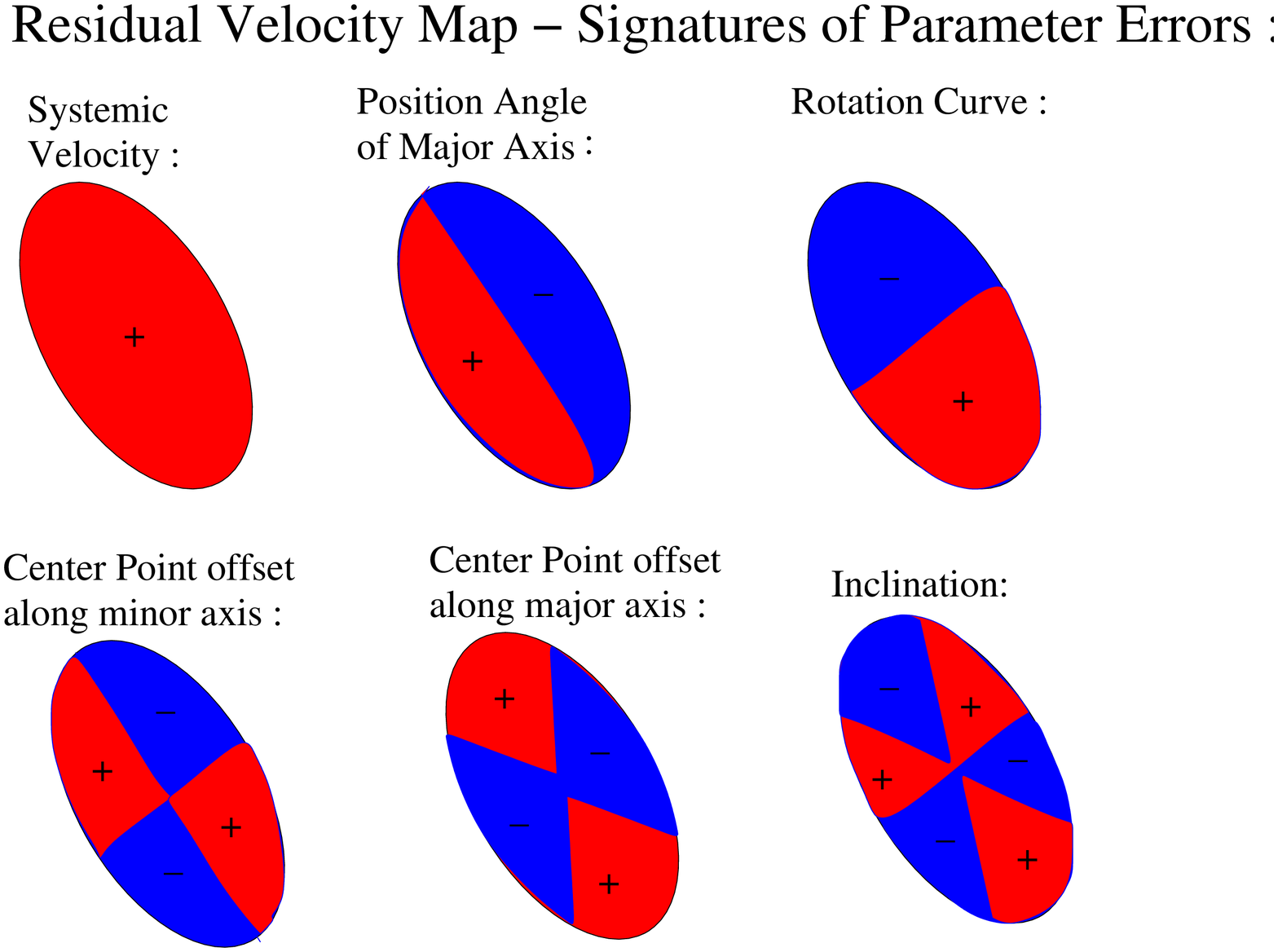}

\noindent {\small Figure 7.  Errors in the fitted parameters
give rise to patterns in the velocity residual map.}

\section{Conclusions}

Single dish telescopes are indispensible for observations of any
spectral line source which is so extended that it cannot be 
imaged in a single interferometer field.  The combination of 
single dish and aperture synthesis data is becoming the standard
technique for mapping Galactic sources, especially for the 
ubiquitous HI line.  Single dish telescopes like Arecibo and 
the GBT will be hugely valuable for this, particularly when
equipped with multi-beam receivers to enhance their mapping 
speed.  We may be on the threshold of a renaissance in single
dish, Galactic, spectral line astronomy.

\end{document}